\documentclass[twocolumn,showpacs,amssymb,aps]{revtex4}
\usepackage{amsmath}
\usepackage{graphicx}
\begin{document}

\title{Hard thermal effective action in QCD through the thermal operator}  

\author{Ashok Das$^{a,b}$ and J. Frenkel$^{c}$}
\affiliation{$^{a}$ Department of Physics and Astronomy,
University of Rochester,
Rochester, NY 14627-0171, USA}
\affiliation{$^{b}$ Saha Institute of Nuclear Physics, 1/AF
  Bidhannagar, Calcutta 700064, INDIA}
\affiliation{$^{c}$ Instituto de F\'{\i}sica, Universidade de S\~ao
Paulo, S\~ao Paulo, SP 05315-970, BRAZIL}

\bigskip

\begin{abstract}

We derive in a simple way the
well known hard thermal effective action for QCD through the
application of the thermal operator to the zero
temperature retarded Green's  functions.
This derivation also clarifies the origin of important properties of
the hard thermal effective action, such as the manifest Lorentz and
gauge invariance of its integrand, by relating them directly to the
properties of the corresponding zero temperature effective action in
the hard regime.
\end{abstract}


\maketitle

\section{Introduction}

The high temperature limit of thermal QCD is of much interest because
of its relevance in the study of the physical properties of the
quark-gluon plasma \cite{kapusta,lebellac}. Several important features
in this regime are embodied in the leading hard thermal effective
action, whose overall coefficient is proportional to $T^{2}$, where
$T$ denotes the equilibrium temperature. Such leading contributions to
the effective action arise at one loop from diagrams where the
internal momentum is hard (being of the order $T$) and is much larger
than any external momenta. The hard thermal effective action in QCD,
which enjoys various symmetry properties (at the integrand level) such
as manifest Lorentz and gauge invariance \cite{braaten,frenkel}, has
also been studied from various points of view
\cite{taylor,jackiw,nair,blaizot,kelley,das1}. 

The purpose of this note is to present a simple derivation of this
action which may explain the origin of these interesting symmetry
properties. The 
derivation is based on an interesting relation between Feynman graphs
at finite temperature and the corresponding zero temperature graphs
\cite{espinosa,silvana}, which holds both in the imaginary time as well as in
the real time formalisms \cite{kapusta,lebellac,das}. This relation,
known as the thermal operator representation, arises as a
consequence of the fact that the thermal propagator for a bosonic
field can be related to
the zero temperature one through a simple thermal operator which
carries the entire temperature dependence and has the explicit form
\begin{equation}
O^{(T)} (E) = 1 + n_{\rm B} (E) (1 - S(E)).\label{tor}
\end{equation}
Here $E = \sqrt{\vec{k}^{\ 2} + m^{2}}, S(E)$ is a reflection operator
that takes $E\rightarrow -E$ and $n_{\rm B} (E)$ represents the
bosonic distribution function. (For a fermionic field a similar
relation holds with $n_{\rm B}\rightarrow -n_{\rm F}$
\cite{silvana}.) This relation between the finite
temperature Feynman graphs and the zero temperature ones is
calculationally quite useful and allows us to study directly many
questions of interest such as the cutting rules at finite temperature 
\cite{silvana1}. Furthermore, the relation between the retarded
thermal Green's functions and the  forward scattering
amplitudes for on-shell thermal particles \cite{brandt} has been
clarified through the application of the thermal operator
representation to the
corresponding zero temperature forward scattering amplitudes
\cite{silvana2}. In this paper, we further demonstrate the simplicity
and the utility of the
thermal operator representation by deriving the hard thermal effective action
for QCD from the zero temperature retarded amplitudes in the hard regime. This
derivation directly associates the origin of the interesting
properties of the hard thermal effective  action, such as Lorentz and
gauge invariance of the integrand, to those of the zero temperature
effective action.

The paper is organized as follows. In section {\bf II}, we show that
in the hard region at zero temperature, the forward scattering
amplitudes are Lorentz and gauge covariant and obey simple Ward
identities. These features, together with the fact that the leading
one loop contributions are quadratic in the hard internal momentum,
are sufficient to determine uniquely all the hard $n$-point gluon
amplitudes in terms of the hard gluon self-energy. An explicit example
of how this works for the 3-point gluon amplitude is discussed in more
detail in the appendix. In section {\bf III}, we are thus able to
write down a generating functional (effective action) for all the hard
$n$-point gluon amplitudes at zero temperature. Through the
application of the thermal operator, it is then immediate to arrive at the
well known form of the hard thermal effective action. In this way, the
symmetry properties of the action, such as Lorentz and gauge
invariance, can be directly understood in terms of the properties of
the zero temperature retarded amplitudes. We conclude with a brief
summary in section {\bf IV}.

\section{Hard forward scattering amplitudes at zero temperature}

The forward scattering amplitude associated with the retarded gluon
self-energy at one loop, $\Pi_{\mu\nu}^{ab} (p)$, can be described as
in Fig. 1,
where ``R" denotes a retarded propagator while the cut line with
momentum $k$ represents an on-shell particle scattering in the forward
direction \cite{silvana2}. In the hard region where the internal momentum $k\gg p$,
with $k^{2}=0$, we can expand the denominator (of the retarded
propagator where $k_{0}$ is to be understood as $k_{0}+i\epsilon$) in
the graph as
\begin{equation}
\frac{1}{(k+p)^{2}} = \frac{1}{2k\cdot p} - \frac{p^{2}}{(2k\cdot
  p)^{2}} + \cdots .\label{expansion} 
\end{equation}
The important thing to note here, for later use, is that this
expansion does not involve any factor of $p^{2}$ in the denominator.

\begin{widetext}
\begin{center}
\begin{figure}[ht!]
\includegraphics[scale=1]{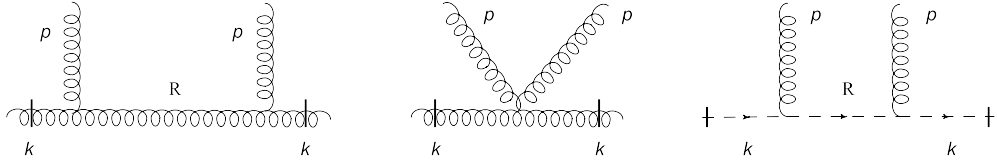}
\caption{Forward scattering graphs for the retarded gluon self-energy
  at one loop. Wavy and dashed lines denote respectively gluon and
  ghost particles. Color and Lorentz indices are suppressed for
  simplicity.}
\end{figure}
\end{center}
\end{widetext}
Similarly, expanding the numerator of the graph in powers of the
external momentum, we find that the leading hard contribution to the
gluon self-energy, $\Pi_{\mu\nu}^{ab} (p)= \delta^{ab} \Pi_{\mu\nu}
(p)$, in the $SU(N)$ Yang-Mills theory can be written as \cite{silvana2}
\begin{widetext}
\begin{equation}
\Pi_{\mu\nu} (p) = 2g^{2}N \int \frac{d^{4}k}{(2\pi)^{3}}\ \delta_{+}
(k^{2})\  \left(\eta_{\mu\nu} -
  \frac{k_{\mu}p_{\nu}+k_{\nu}p_{\mu}}{k\cdot p} + p^{2}
  \frac{k_{\mu}k_{\nu}}{(k\cdot p)^{2}}\right),\label{selfenergy}
\end{equation}
\end{widetext}
where we have defined
\begin{equation}
\delta_{+} (k^{2}) = \theta (k_{0}) \delta (k^{2}).
\end{equation}

There are several things to note from the structure of the gluon
self-energy in \eqref{selfenergy}. The quantity in the parenthesis is
manifestly Lorentz covariant being of zero degree in the internal
momentum. Consequently, the integrand is quadratic in $k$ for large
values of the internal momentum and, as we will show shortly, this
leads to the leading contribution of order $T^{2}$ at high
temperature. To apply the thermal operator, we need to carry out the
integration over $k_{0}$ \cite{silvana} in which case
\eqref{selfenergy} takes the form
\begin{widetext}
\begin{equation}
\Pi_{\mu\nu} (p) = 2g^{2}N \int
\frac{d^{3}k}{(2\pi)^{3}}\frac{1}{2E_{k}}\ \left(\eta_{\mu\nu} -
  \frac{\hat{k}_{\mu}p_{\nu}+\hat{k}_{\nu}p_{\mu}}{\hat{k}\cdot p} +
  p^{2} \frac{\hat{k}_{\mu}\hat{k}_{\nu}}{(\hat{k}\cdot
    p)^{2}}\right),\label{selfenergy1}
\end{equation}
\end{widetext}
where $E_{k}= |\vec{k}|$ and we have defined $\hat{k}_{\mu} = (1, -
\hat{\vec{k}})$. (We use a metric with signature $(+,-,-,-)$.) It is
worth remarking here that the self-energy in \eqref{selfenergy1} is
gauge independent and satisfies the Ward identity
\begin{equation}
p^{\mu} \Pi_{\mu\nu} (p) = 0.\label{transversality}
\end{equation}

Next, let us consider the retarded $n$-point gluon forward scattering
amplitude at one loop. As discussed in \cite{silvana2}, this is
described by graphs with a single on-shell cut propagator, all other
propagators in a graph being retarded/advanced. Thus, one can write
such an amplitude in the form (compare with \eqref{selfenergy})
\begin{widetext}
\begin{equation}
\Gamma_{\mu_{1}\mu_{2}\cdots \mu_{n}}^{a_{1}a_{2}\cdots a_{n}}
(p_{1},p_{2}\cdots ,p_{n}) = \int\frac{d^{4}k}{(2\pi)^{3}}\ \delta_{+}
(k^{2})\ \gamma_{\mu_{1}\mu_{2}\cdots \mu_{n}}^{a_{1}a_{2}\cdots
  a_{n}} (k,p_{1},\cdots , p_{n}),\label{npoint}
\end{equation}
\end{widetext}
where $\gamma$ represents the forward scattering amplitude for an
on-shell particle of momentum $k$. For large values of $k$, we can
again expand the denominators in $\gamma$ (see \eqref{expansion}) in
powers of $p^{2}/(2k\cdot p)$, where $p$ denotes a linear combination
of the external momenta. Furthermore, we can also expand the numerator
in powers of $p_{i,\mu}/|\vec{k}|$. The first term in the expansion
comes from the leading term in each of the retarded propagators of the
form $1/(k\cdot p)$ and a numerator independent of $p_{i}$. The
contributions to $\gamma$ from these superleading terms, which are of
degree one in $k$, however vanish by symmetry under $k\rightarrow
-k$. The next term in the expansion which does provide the leading
quadratic contribution to the $n$-point gluon amplitude comes from
terms in $\gamma$ which are functions of degree zero in $k$
\cite{brandt}.

In the hard regime, to leading order, these amplitudes obey simple
linear Ward identities. This follows as a consequence of the fact that
diagrams with external (open) ghost lines, which appear in the BRST
identities, have one less power of $k$ in the numerator compared with
the corresponding diagrams involving only external gluon lines. As a
result, the first term in the expansion of such a graphs is also
naively  quadratic in $k$ much like the leading contribution for the
gluon amplitude. However, in the case of graphs with external ghost
lines, these leading contributions cancel by an eikonal identity when
all graphs are added. Therefore, the contribution of the amplitude
becomes subleading compared to the gluon amplitude. In the appendix,
we illustrate how such a cancellation takes place for the one loop
ghost-ghost-gluon (3-point) amplitude. As a consequence, the leading
terms in the $n$-point gluon amplitude obey simple Ward
identities. For example, to leading order, the 3-point gluon amplitude
satisfies the relation
\begin{equation}
p_{3}^{\lambda} \Gamma_{\mu\nu\lambda}^{abc} (p_{1},p_{2},p_{3}) = ig
f^{abc}\left[\Pi_{\mu\nu} (p_{1}) - \Pi_{\mu\nu}
  (p_{2})\right],\label{ward}
\end{equation}
showing that it is related to the hard self-energy. Similarly, the
Ward identities relate the 4-point gluon amplitude to the 3-point
amplitude as
\begin{widetext}
\begin{equation}
p_{4}^{\rho}\Gamma_{\mu\nu\lambda\rho}^{abcd}
(p_{1},p_{2},p_{3},p_{4}) = ig\left[f^{cde}
  \Gamma_{\mu\nu\lambda}^{abe} (p_{1},p_{2},p_{3}+p_{4}) +
  f^{bde}\Gamma_{\mu\nu\lambda}^{cae} (p_{1},p_{2}+p_{4},p_{3}) +
  f^{ade}\Gamma_{\mu\nu\lambda}^{bce}
  (p_{1}+p_{4},p_{2},p_{3})\right].\label{ward1}
\end{equation}
\end{widetext}
The above properties are sufficient to determine uniquely to leading
order, all the $n$-point gluon amplitudes in terms of the hard gluon
self-energy at zero temperature. This is discussed in detail in the
appendix for the 3-point gluon amplitude, but similar arguments hold
for all $n$-point gluon amplitudes. (We would like to emphasize here
that, in general, the solution to the Ward identities such as
\eqref{ward} is not unique since the transverse part of the amplitude
is not fully determined by the identity. However, in the hard region,
the fact that the leading term has an integrand of zero degree in $k$
and that the denominators have the form $k\cdot p$ as in
\eqref{expansion} is sufficient to determine all the amplitudes
uniquely.)

\section{The hard effective action and its properties}

The simple Ward identities satisfied by the hard $n$-point gluon
amplitudes can be written in a compact form as
\begin{equation}
D_{\mu}^{ab} \frac{\delta \Gamma[A]}{\delta A_{\mu}^{b} (x)} =
\left(\delta^{ab}\partial_{\mu} - gf^{abc} A_{\mu}^{c}\right)
\frac{\delta\Gamma[A]}{\delta A_{\mu}^{b} (x)} =
0,\label{wardidentity}
\end{equation}
where $\Gamma[A]$ denotes the generating functional (effective action)
and $D_{\mu}^{ab}$ is the covariant derivative. Relation
\eqref{wardidentity} implies that the effective action, $\Gamma[A]$,
is invariant under an infinitesimal non-Abelian gauge transformation
with parameter $\omega^{a} (x)$, namely, under 
\begin{equation}
A_{\mu}^{a}\rightarrow A_{\mu}^{a (\omega)} = A_{\mu}^{a} +
D_{\mu}^{ab}\omega^{b},\quad \Gamma[A]\rightarrow
\Gamma[A^{(\omega)}],
\end{equation}
the infinitesimal change in the effective action is given by
\begin{equation}
\left.\frac{\delta\Gamma[A^{(\omega)}]}{\delta
    \omega^{a}}\right|_{\omega=0} = \frac{\delta A_{\mu}^{b
    (\omega)}}{\delta \omega^{a}} \frac{\delta\Gamma[A]}{\delta
  A_{\mu}^{b}} = - D_{\mu}^{ab} \frac{\delta\Gamma[A]}{\delta
  A_{\mu}^{b}} = 0,
\end{equation}
where we have used a compact notation suppressing all the intermediate
integrations. 

Using \eqref{npoint} and performing the $k_{0}$ integration, we can
write the effective action to leading order in the form 
\begin{equation}
\Gamma[A] = \int d^{4}x \int
\frac{d^{3}k}{(2\pi)^{3}}\frac{1}{2E_{k}}\ \gamma (A, k).
\end{equation}
From our earlier discussion, we note that $\gamma$ has the following
characteristics.
\begin{enumerate}
\item It is gauge invariant.
\item It is a Lorentz invariant function of zero degree in $k_{\mu}$
  with $k_{0}=|\vec{k}|$.
\item The integrand involves denominators with products of factors of
  the form $k\cdot p$ where $p$ is some linear combination of external
  momenta.
\end{enumerate} 
As we have discussed earlier, in the hard region, all the $n$-point
gluon amplitudes are determined uniquely from the hard gluon
self-energy. Therefore, any effective action satisfying the properties
listed above and yielding the correct one loop hard gluon self-energy
would correspond to the unique hard effective action.  Comparing with
the hard gluon self-energy in \eqref{selfenergy1}, it follows that the
hard effective action at zero temperature can be written in the form
\begin{equation}
\Gamma[A] = \frac{g^{2}N}{(2\pi)^{3}} \int d^{4}x \int
\frac{d^{3}k}{2E_{k}}\ {\cal A}_{\mu}^{a} (x,\hat{k}) {\cal A}^{\mu a}
(x, \hat{k}),\label{zeroT}
\end{equation}
where the gauge covariant potential, introduced in \cite{das1}, has the form
\begin{eqnarray}
{\cal A}_{\mu}^{a} (x,\hat{k}) &=& \left(\frac{1}{\hat{k}\cdot D}
  \hat{k}^{\nu} F_{\nu\mu}\right)^{a}\nonumber\\
&=& A_{\mu}^{a} - \left(\frac{1}{\hat{k}\cdot D} \partial_{\mu}
  \hat{k}\cdot A\right)^{a},\label{covariantpotential}
\end{eqnarray}
and $F_{\mu\nu}^{a}$ denotes the non-Abelian field strength tensor
\begin{equation}
F_{\mu\nu}^{a} = \partial_{\mu}A_{\nu}^{a} - \partial_{\nu}A_{\mu}^{a}
+ gf^{abc} A_{\mu}^{b}A_{\nu}^{c}.
\end{equation}
The gauge covariant non-Abelian potential, ${\cal A}_{\mu}^{a}
(x,\hat{k})$, is in general a nonlocal function, where the nonlocality
arises from retarded line integrals along the direction of $\hat{k}$.

Once the hard effective action \eqref{zeroT} is determined at zero
temperature, the hard thermal effective action can be determined by
applying the thermal operator in the following manner. It is worth
recalling \cite{silvana,silvana2} that the thermal operator acts on
functions of energy in the integrand before the spatial momentum
integrations are carried out. Furthermore, in spite of the fact that
the perturbative expansion of the covariant potential in
\eqref{covariantpotential} involves retarded propagators, the thermal
operator leaves such propagators invariant \cite{silvana2} so that the
covariant gauge potential is unaffected by the application of the
thermal operator. As a consequence, the only term in the integrand on
which the thermal operator  \eqref{tor} acts nontrivially is the
energy denominator (coming from the on-shell cut propagator)
\begin{equation}
O^{(T)} (E_{k}) \frac{1}{2E_{k}} = \frac{1}{2E_{k}}\left(1 + 2 n_{\rm
    B} (E_{k})\right).
\end{equation}
Thus, the application of the thermal operator immediately leads to the
temperature dependent hard thermal effective action as
\begin{equation}
\Gamma^{(\beta)}[A] = \frac{g^{2}N}{(2\pi)^{3}} \int d^{4}x \int
d^{3}k\ \frac{n_{\rm B} (E_{k})}{E_{k}}\ {\cal A}_{\mu}^{a}
(x,\hat{k}) {\cal A}^{\mu a} (x,\hat{k}).
\end{equation}
Since $E_{k} = |\vec{k}|$, the radial momentum integration can be
carried out using the standard integral (with the Boltzmann constant
set to unity)
\begin{equation}
\int_{0}^{\infty} dk\ kn_{\rm B} (E_{k}) = \int_{0}^{\infty} dk\
\frac{k}{e^{k/T} - 1} = \frac{\pi^{2}T^{2}}{6},
\end{equation}
which yields the hard thermal effective action
\begin{equation}
\Gamma^{(\beta)}[A] = \frac{g^{2}T^{2}N}{12} \int d^{4}x \int
\frac{d\Omega}{4\pi} {\cal A}_{\mu}^{a} (x,\hat{k}){\cal A}^{\mu a}
(x,\hat{k}).
\end{equation}
Here $d\Omega$ represents the angular integration over the unit vector
$\hat{\vec{k}}$. Finally, integrating this expression by parts, we can
recast it in the well known form of the hard thermal effective action
\cite{braaten,frenkel}
\begin{equation}
\Gamma^{(\beta)}[A] = \frac{m_{\rm gl}^{2}}{2} \int d^{4}x \int
\frac{d\Omega}{4\pi}\ F^{\mu\nu a}
\left(\frac{\hat{k}_{\nu}\hat{k}^{\lambda}}{\left(\hat{k}\cdot
      D\right)^{2}}\right)^{ab} F_{\lambda\mu}^{b},\label{hardT}
\end{equation}
where we have identified $m_{\rm gl}^{2} = g^{2}T^{2}N/6$ as the
square of the thermal gluon mass.

It is now straightforward to extend the above result for the pure
Yang-Mills theory to QCD with quarks in the fundamental
representation.  To this end, we recall that the thermal
operator representation works also for theories involving fermions
\cite{silvana}. In this case, as we have pointed out earlier, the thermal 
operator, relating the propagator at zero temperature to that at finite
temperature, has a  form similar to the one 
given in \eqref{tor}, with $n_{\rm B}(E)\rightarrow - n_{\rm F} (E)$,
where $n_{\rm F}$ denotes the fermionic distribution function. Because
of the linearity of the Ward identities in the hard regime, the
contribution of the hard quark loops to an amplitude, which is
additive, independently satisfies simple Ward identities like those 
given in Eqs. \eqref{ward} and \eqref{ward1} \cite{frenkel}. As shown
in the appendix, these identities together with the fact that the
leading order terms in the integrand are Lorentz covariant functions
of degree zero in the internal loop momentum are sufficient to
determine uniquely all the higher $n$-point gluon amplitudes in the
hard regime in terms of the corresponding gluon
self-energy. Given the above properties of hard amplitudes as
well as the transversality condition in \eqref{transversality}, it follows
that the integrand for the self-energy with a hard quark loop
at zero temperature necessarily has a
structure similar to the one given in \eqref{selfenergy}, upto an overall
multiplicative factor which can be calculated easily. Consequently,
adding this contribution to that of the pure Yang-Mills theory in
\eqref{selfenergy}, the hard thermal
effective action for QCD, which can be obtained by applying the
thermal operator, 
has a similar form to the one given in \eqref{hardT} with the square
of the thermal gluon mass given by
\begin{equation}
m_{\rm QCD}^{2} = \frac{g^{2}T^{2}}{6}\left(N + \frac{1}{2}
  N_{\rm f}\right),
\end{equation}
where $N_{\rm f}$ denotes the number of quark flavors.

\section{Conclusion}

In this work, we have used the forward scattering description for the
retarded amplitudes in QCD to construct the hard effective action at
one loop at zero temperature. By applying the thermal operator to the
zero temperature amplitudes, we have derived in a simple way the hard
thermal effective action for QCD. This approach emphasizes that
various relevant features of this action such as Lorentz invariance of
the integrand in \eqref{hardT} as well as its gauge invariance, simply
arise because the leading hard zero temperature amplitudes (and the
effective action) precisely have such properties. (The angular
integration breaks Lorentz invariance at finite temperature, as
expected, since the rest frame of the heat bath defines a preferred
reference frame. An alternative form with a Lorentz non-invariant
integrand has been given in \cite{taylor}.) The above properties of
the hard thermal effective action turn out to be very convenient in
the study of the energy-momentum tensor for the quark-gluon plasma, as
well as in the analysis of the high temperature behavior of gauge
field theories in a curved space-time \cite{frenkel1}. Furthermore,
the effective action \eqref{hardT} is also useful in implementing the
resummation procedure \cite{braaten} which is necessary for a
consistent perturbative expansion in hard thermal QCD.

\vskip 1cm

\noindent{\bf Acknowledgment:}
\medskip

One of us (AD) acknowledges the Fulbright Foundation for a
fellowship. This work was
supported in part by US DOE Grant number DE-FG 02-91ER40685, by CNPq
and by FAPESP, Brazil.

\appendix*

\section{Hard amplitudes from Ward identities}

For definiteness, we discuss here in more detail some of the
properties of the hard 3-point amplitudes at one loop. First, let us
look at the ghost-ghost-gluon vertex, $V_{\mu}^{abc}
(p_{1},p_{2},p_{3})$ at one loop and show how the leading terms cancel
in this case. Let us examine the forward scattering graphs associated
with such a vertex correction at one loop shown in Fig. 2, where
$p_{1}+p_{2}+p_{3} = 0$.
\begin{widetext}
\begin{center}
\begin{figure}[ht!]
\includegraphics[scale=.6]{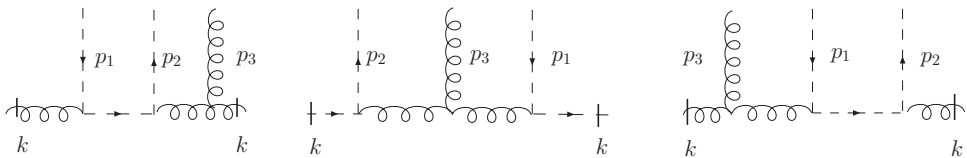}
\vspace{-.5in}
\caption{Examples of forward scattering graphs associated with the
  retarded ghost-ghost-gluon vertex at one loop. Color and Lorentz
  indices are suppressed for simplicity.}
\end{figure}
\end{center}
\end{widetext}

In the hard regime (apart from an overall color factor $f^{abc}$)
these graphs have a common numerator $(p_{2}\cdot k)k_{\mu}$. The
presence of quadratic terms in $k$ in the numerator means that the
leading terms, as in \eqref{expansion}, of the two retarded
propagators (in the graph with $k^{2}=0$) are sufficient to give us
the leading terms of degree zero in $k$. However, when the three
graphs are summed, this contribution cancels as a consequence of the
eikonal identity
\begin{equation}
\frac{1}{(k\cdot p_{1})(k\cdot p_{2})} + \frac{1}{(k\cdot
  p_{2})(k\cdot p_{3})} + \frac{1}{(k\cdot p_{3})(k\cdot p_{1})} = 0.
\end{equation}
Therefore, the ghost-ghost-gluon amplitude has a subleading
contribution compared with the three point gluon amplitude, which as a
result, satisfies the simpler Ward identity \eqref{ward}.

Let us next show that the Ward identity \eqref{ward}, together with
the fact that to leading order the integrand of the 3-point gluon
amplitude, $\gamma_{\mu\nu\lambda}^{abc} (k,p_{1},p_{2},p_{3})$, is a
function of zero degree in $k$, is sufficient to determine the hard
3-point gluon amplitude in terms of the hard gluon self-energy in
\eqref{selfenergy}. We note that if we factor out an overall color
factor of $f^{abc}$, the Lorentz structure of the leading terms in
$\gamma_{\mu\nu\lambda}^{abc}$ (which are of degree zero in $k$ and
have the dimension of an inverse mass) can be parameterized in general
as
\begin{eqnarray}
& & B k_{\mu}k_{\nu}k_{\lambda} + \sum_{i} \left(C_{1 i}p_{i
    \mu}k_{\nu}k_{\lambda} + C_{2 i}k_{\mu}p_{i \nu} k_{\lambda} +
  C_{3 i} k_{\mu}k_{\nu}p_{i\lambda}\right)\nonumber\\ 
& &\quad + \left(E_{1} k_{\mu}\eta_{\nu\lambda} + E_{2} k_{\nu}
  \eta_{\mu\lambda} +  E_{3} k_{\lambda}
  \eta_{\mu\nu}\right),\label{lorentz} 
\end{eqnarray}
where the coefficient functions $B, C_{1 i}, C_{2 i}, C_{3 i}, E_{1},
E_{2}, E_{3}$ are Lorentz invariant functions of $k$ and $p_{i}$. The
important point to note from \eqref{lorentz} is that it is at most
linear in $p_{i}$ as well as in the metric tensor. (It is worth
remarking here that while a Lorentz structure such as $D_{ij}p_{i \mu}
p_{j \nu} k_{\lambda}$ is allowed in principle, for such a term to be
of degree zero in $k$ and have the inverse dimension of mass, the
coefficient $D_{ij}$ must contain a denominator of the form  $p^{2}$
where $p$ denotes a linear combination of the external
momenta. However, as we have pointed out earlier, such terms do not
arise in the expansion \eqref{expansion} of the retarded
propagator. This is why the numerator can at most be linear in the
external momenta.)

Given the general Lorentz structure \eqref{lorentz}, let us next
consider the Ward identity \eqref{ward} by contracting the amplitude
with $p_{3}^{\lambda}$ which leads in the integrand to the Lorentz
structure 
\begin{eqnarray}
& & \left(B k\cdot p_{3} + \sum_{i} C_{3i} p_{i}\cdot p_{3}\right)
k_{\mu}k_{\nu} \nonumber\\
& & + \sum_{i} \left(C_{1 i} k\cdot p_{3} + E_{2} \delta_{3i}\right)
p_{i \mu}k_{\nu} \nonumber\\
& & + \sum_{i}\left(C_{2 i} k\cdot p_{3} + E_{1} \delta_{3i}\right)
p_{i \nu} k_{\mu} + E_{3} k\cdot p_{3} \eta_{\mu\nu}.\label{structure}
\end{eqnarray}
We can now compare this structure to the ones on the right hand side
of the Ward identity in \eqref{ward} coming from the structure of the
self-energy in \eqref{selfenergy}. Comparing the coefficients of the
metric tensor on both sides, the last term in \eqref{structure},
namely, $E_{3}$ is uniquely determined. Likewise, the Ward identity
\eqref{ward} where we contract with $p_{1}^{\mu}$ or $p_{2}^{\nu}$
determines respectively $E_{1}$ and $E_{2}$ uniquely. Similarly,
matching the second and the third structures in \eqref{structure} with
the corresponding structures coming from the self-energy (along with
the other two identities) uniquely determines $\sum_{i} C_{1 i} p_{i
  \mu}, \sum_{i} C_{2 i} p_{i \nu}, \sum_{i} C_{3 i} p_{i
  \lambda}$. Finally, comparing the $k_{\mu}k_{\nu}$ terms on both
sides of the Ward identity \eqref{ward} then uniquely determines
$B$. In this way, the leading  3-point gluon amplitude is uniquely
determined in terms of the hard gluon self-energy. One can follow this
argument to show that the leading higher point gluon amplitudes are
recursively determined uniquely by the hard gluon self-energy. This
fact allows us to determine the effective action for gluon amplitudes
in the hard region in a unique form.

\end{document}